\documentclass[letterpaper,12pt]{article}
\usepackage{amssymb,amsmath}
\usepackage[numbers]{natbib}
\usepackage{color}
\usepackage{txfonts}
\usepackage{multirow}
\usepackage{hyperref}
\usepackage{hypernat}
\usepackage{verbatim}
\usepackage{stmaryrd}
\usepackage{graphicx,float}
\usepackage[affil-it]{authblk}
\usepackage{subcaption}
\usepackage{soul}
\newcommand{\dd}{{\rm d}}
\usepackage{wasysym}
\begin{document}
\renewcommand*{\Authfont}{\normalsize}
\title{Stiff Matter Solution in Brans-Dicke Theory and The General Relativity Limit}

\author[1]{G.~Brando \footnote{Email: \href{mailto:gbrando@cosmo-ufes.org}{gbrando@cosmo-ufes.org}}}
\affil[1]{PPGCosmo, CCE - Universidade Federal do Esp\'\i rito Santo, zip 29075-910, Vit\'oria, ES, Brazil }

\author[1,2,3]{J.~C.~Fabris \footnote{Email: \href{mailto:julio.fabris@cosmo-ufes.org}{julio.fabris@cosmo-ufes.org}}}

\affil[2]{N\'ucleo Cosmo-ufes \& Departamento de F\'\i sica - Universidade Federal do Esp\'\i rito Santo, zip 29075-910, Vit\'oria, ES, Brazil} 
\affil[3]{National  Research  Nuclear  University  MEPhI, Kashirskoe  sh. 31,  Moscow  115409,  Russia} 

\author[1,4]{F.~T.~Falciano \footnote{Email: \href{mailto:ftovar@cbpf.br}{ftovar@cbpf.br}}}

\affil[4]{CBPF - Centro Brasileiro de Pesquisas F\'{\i}sicas, Xavier Sigaud st. 150,	zip 22290-180, Rio de Janeiro, Brazil} 

\author[5]{Olesya Galkina \footnote{Email: \href{mailto:ole.galkina@gmail.comr}{ole.galkina@gmail.com}}}

\affil[5]{PPGFis, CCE - Universidade Federal do Esp\'\i rito Santo, zip 29075-910, Vit\'oria, ES, Brazil}

\date{}

\maketitle
\begin{abstract}
Generally the Brans-Dicke theory reduces to General Relativity in the limit $\omega\rightarrow\infty$ if the scalar field goes as $\phi\propto1/\omega$. However, it is also known that there are examples with $\phi\propto1/\sqrt{\omega}$ that does not tend to GR. We discuss another case: a homogeneous and isotropic universe filled with stiff matter. The power of time dependence of these solutions do not depend on $\omega$, and there is no General Relativity limit even though we have $\phi\propto1/\omega$. A perturbative and a dynamical system analysis of this exotic case are carried out.
\end{abstract}

\maketitle

\section{Introduction}
General Relativity (GR) is a very successful theory to describe gravitational interaction. It is a theoretically consistent and experimentally tested theory. So far, GR has been confirmed by every experiment in the solar system~\cite{will1993, reasenberg1973, everitt2011,rodrigues2018} and astrophysical phenomena such as the emission of gravitational waves by binary systems and it is in accordance with the bounds on the velocity of gravitational waves \cite{weisberg2010, abbott2017}. However, there are at least three reasons to seriously consider alternative theories of gravity. The necessity of the dark sector (dark energy and dark matter) on the standard cosmological model; the theoretical motivation to unify the gravitational interaction with the quantum interactions in a single theoretical framework; and the epistemological fact that alternative theories can be used to highlight the intrinsic properties of GR by showing how it could be otherwise. 

The prototype of alternative theory of gravity is Brans-Dicke theory (BD). Historically, it is one of the most important alternative to the standard General Relativity theory, which was introduced by C. Brans and R. H. Dicke \cite{brans1961} as a possible implementation of Mach's principle in a relativistic theory. 
Solar System time-delay experiments set a lower bound on the absolute value of the dimensionless parameter $|\omega| > 500$~\cite{will1993, reasenberg1973}, which means that BD is strongly constrained for the solar system dynamics. The theory is also constrained by the CMB, as pointed out in~\cite{wu2010,avilez2014}. Notwithstanding, there is phenomenological applications for BD in cosmology and indeed it has received recently much attention of the scientific community~\cite{alexander2016,tretyakova2015,hrycyna2013,will2004,kofinas2016,papagiannopoulos2016,alonso2016,roy2017,gerard1995,barrow2008,brando2018}.

In the present work, we display a particular solution of BD with matter content described by a stiff matter barotropic perfect fluid. This is a very interesting solution with exotic characteristics revealing some of the new features, for better or worst, that one can expect to find in BD-like alternative theories of gravity~\cite{stoycho,holdenwands,barrowparsons,faraoni2009}. In particular, the time evolution of the system is independent of the value of the parameter $\omega$. The evolution of the perturbations has only growing modes, which is also another distinct feature of this solution. In addition, the scale factor evolution behaves as $a\propto t^{1/2}$, typical of radiation dominated epoch in GR, hence this configuration might have some application in the early universe. Recentely, this period of the universe filled with stiff matter was also studied in the context of $f(R)$ theories~\cite{odintsovoikonomou}.

It is argued in the literature that BD approaches GR in the $|\omega|\rightarrow \infty$ limit~\cite{weinberg1972}. The crucial point behind this argument is that when the parameter $|\omega| \gg 1$, the field equations seem to show that $\square \varphi=\mathcal{O}\left(\frac1\omega\right)$ and hence 
\begin{align}
\varphi=&\frac{1}{G_N} +\mathcal{O}\left(\frac{1}{\omega}\right)	\label{BDeq12}\\
G_{\mu\nu}=&8\pi G_N T_{\mu\nu}+\mathcal{O}\left(\frac{1}{\omega}\right)\label{BDeq22}
\end{align}
where $G_N$ is Newton's gravitational constant, $G_{\mu\nu}$ is the Einstein tensor and assume natural unit where $c=1$. However, there are some examples~\cite{anchordoqui1998,jarv2015,nariai1968,banerjee1985,banerjee1986,banerjee1997,hanlon1972,matsuda1972,romero1993a,romero1998,romero1993b,paiva1993a,paiva1993b,scheel1995,faraoni1999,chauvineau2003} where exact solutions can not be continuously deformed into the corresponding\footnote{the word corresponding here is used in the sense of the same matter content as in GR.} GR solutions by taking the  $|\omega|\rightarrow\infty$ limit. Their asymptotic behavior differs exactly because these solutions do not decay as Eq.~\eqref{BDeq12} but instead as
\begin{equation}\label{BDeq13}
\varphi=\frac1{G_N} +\mathcal{O}\left(\frac{1}{\sqrt{\omega}}\right)\quad .
\end{equation}

Our particular solution, to be developed in section~\ref{sec:PartSol}, has the novelty of having the appropriate asymptotic behavior given by Eq.~\eqref{BDeq12} (see Eq.~\eqref{phi0rho0wrel}) but no GR limit.

The paper is organized as follows. In section~\ref{sec:EqMot} we briefly describe the system and its equation of motion. In section~\ref{sec:GureSol} we review the general solution studied by Gurevich et al. In section~\ref{sec:PartSol} we analyze a power law solution with peculiar features and develop the perturbation over this specific background in section~\ref{sec:Pert}. In section~\eqref{sec:DynSys} we perform a dynamical system analysis of the system and finally in section~\ref{sec:Concl} we end with some final remarks.

\section{The classical equations of motion}\label{sec:EqMot}

In Brans-Dicke theory the scalar field is understood as part of the geometrical degrees of freedom. This theory has a non-minimal coupling between gravity and the  scalar field. The action reads
\begin{equation}\label{eq:Lagrangian}
S=\int \dd x^4 \sqrt{-g}\left[\frac1{16\pi}\left(\phi R-\frac{\omega}{\phi}\nabla_{\alpha}\phi\nabla^{\alpha}\phi
\right)+\mathcal{L}_{\mathrm{m}}\right] \ ,
\end{equation}
where $\mathcal{L}_{m}$ is the ordinary matter and $\omega$ is the scalar field coupling constant. Variation of the action Eq.~\eqref{eq:Lagrangian} with respect to the metric and the scalar field give, respectively, the following field equations
\begin{align}
& G_{\mu \nu} = \frac{8 \pi}{\phi} T_{\mu \nu} + \frac{\omega}{\phi^{2}} \left( \nabla_{\mu} \phi \nabla_{\nu} \phi - \frac{1}{2} g_{\mu \nu} \nabla^{\alpha} \phi \nabla_{\alpha} \phi \right) +\frac{1}{\phi} \left( \nabla_{\mu} \nabla_{\nu} \phi - g_{\mu \nu} \Box \phi  \right), \label{eq:fe-1}\\
&\Box \phi = - \frac{\phi}{2 \omega} R +\frac{1}{2\phi} \left( \nabla^{\alpha} \phi \nabla_{\alpha} \phi \right)=\frac{8\pi}{3+2\omega}T \ , \label{eq:fe-2}
\end{align}
where we have used the trace of Eq.~\eqref{eq:fe-1} in the last step of Eq.~\eqref{eq:fe-2}. The action Eq.~\eqref{eq:Lagrangian} is diffeomorphic invariant and since all variables are dynamic fields we have conservation of energy-momentum, i.e.
\begin{align}\label{conservTmunu}
\nabla_\mu T^{\mu \nu}=0 \ .
\end{align}

We shall consider the matter content described by a perfect fluid such that the energy-momentum tensor is
\begin{equation}
T^{\mu\nu}=\left(\rho+p\right)u^{\mu}u^{\nu}-pg^{\mu\nu}\ ,
\end{equation}
and a barotropic equation of state $p=\alpha\rho$ with $,0\leq\alpha\leq 1$. The equation of state parameter $\alpha$ is bounded from above in order to avoid superluminal  speed of sound. In the extreme case, $\alpha=1$, the speed of sound equals the speed of light, which corresponds to stiff matter. This equation of state was first proposed by Zeldovich as an attempt to describe matter in extremely dense states such as in the very early universe.

We shall restrict our analysis to the Friedmann-Lema\^\i tre-Robertson-Walker (FLRW) universes where the metric has a preferred foliation given by homogeneous and isotropic spatial sections. In spherical coordinate system for this particular foliation, the line element has the form
\begin{equation}\label{flrwmetric}
\dd s^2=\dd t^2-a^2(t)\left[\frac{\dd r^2}{1-kr^2}+r^2\left(\dd \theta^2+\sin^2(\theta)\, \dd \phi\right)\right] \ ,
\end{equation}
where $a(t)$ is the scale factor function and $k=0,\pm 1$ defines the spatial section curvature. 

A stiff matter fluid has equation of state $p=\rho$. Thus, conservation of the energy-momentum tensor in a FLRW universe implies  $\rho=\rho_{0}\, (a_0/a)^{6}$ with $\rho_{0}$ and $a_0$ two constants of integration. The flat FLRW case has been analyzed in a quite general form in Ref. \cite{Gurevich} but not all solutions were fully explored. In particular, we describe some exotic features of a peculiar solution associated with this equation of state. After integrating the conservation of energy-momentum equation, the two remaining independent equations become
\begin{align}
\left(\frac{\dot{a}}{a}\right)^{2} +\frac{k}{a^2}& =  \frac{8\pi\rho_{0}}{3\phi}\left(\frac{a_0}{a}\right)^{6}+\frac{\omega}{6}\left(\frac{\dot{\phi}}{\phi}\right)^{2}-\frac{\dot{a}}{a}\frac{\dot{\phi}}{\phi}\ , \label{FLRW1}\\
\ddot{\phi}+3\frac{\dot{a}}{a}\dot{\phi} & =  -\frac{16\pi\rho_{0}}{\left(3+2\omega\right)}\left(\frac{a_0}{a}\right)^{6}\ ,\label{KG1}
\end{align}
where a dot denotes differentiation with respect to the cosmic time $t$.

\section{Gurevich's  Families of  Solutions}\label{sec:GureSol}

In this section we present the general solutions of scale factor and the scalar field for the FLRW flat case in the Brans-Dicke theory obtained by Gurevich \emph{et al. }\cite{Gurevich}. We shall follow closely their presentation but adapting specifically for the stiff matter case. Gurevich \emph{et al.} obtained a class of flat space solutions for
the equation of state $P=\alpha\rho,$ where $0\leq\alpha\leq1$. There are three families of solutions depending on the value of $\Delta\equiv B^2-4AC_0$ where $2A\equiv2(2-3\alpha)+3\omega\left(1-\alpha\right)^2$, $B=3\sigma(1-\alpha)+(1-3\alpha)\beta_0$ and $C_0$ and $\beta_0$ are integration constants. Using the time-time component of Eq.~\eqref{eq:fe-1} and the time component of the conservation of the energy-momentum Eq.~\eqref{conservTmunu}, one can show that for $\alpha\neq 1/3$ the $\Delta$ can be recast as 
\[
\Delta= \frac{(1-3\alpha)^2\sigma^2}{1+2\omega/3}(\beta_0-1)^2 \quad \mbox{with}\quad  \sigma\equiv 1+\omega\left(1-\alpha\right)\quad .
\]
Note that for $\beta_0\neq 1$ the sign of $\Delta$ is negative for $\omega<-3/2$ and positive for $\omega>-3/2$. It is also useful to define a parametric time $\theta$, which is connected to cosmic time through the relation $dt=a^{3\alpha}d\theta.$

The first family of solutions is given by $\Delta<0$ $\left(\omega<-\frac{3}{2}\right)$. The general solutions for the scale factor and scalar field $\phi$ read
\begin{align}
a & =a_{0}\Big[\left(\theta+\theta_{-}\right)^{2}+\theta_{+}^{2}\Big]^{\sigma/{2A}}e^{ \pm \sqrt{\left(\frac{2|\omega|}{3}-1\right)}f(\theta)}\ ,\label{eq:general solution}\\
\phi & =\phi_{0}\left[\left(\theta+\theta_{-}\right)^{2}+\theta_{+}^{2}\right]^{{(1-3\alpha)}/{2A}}e^{ \mp3(1-\alpha) \sqrt{\left(\frac{2|\omega|}{3}-1\right)}f(\theta)}\ , \label{eq:general solution phi}
\end{align}
where
\begin{align}
f(\theta)&= \frac1A\arctan\left(\frac{\theta+\theta_{-}}{\theta_{+}}\right)\ , \quad \theta_{+}=\frac{\sqrt{|\Delta|}}{2A}\ , \quad \theta_{-}=\frac{B}{2A}\quad .\label{eq:general solution A}
\end{align}
For the stiff matter case $(\alpha=1)$, the solutions \eqref{eq:general solution}-\eqref{eq:general solution phi} simplify to

\begin{align}
a&=a_{0}\left[\left(\theta+\theta_{-}\right)^{2}+\theta_{+}^{2}\right]^{-\frac{1}{2}}\exp\left\{\mp\sqrt{\frac{2|\omega|}{3}-1}\arctan\left(\frac{\theta+\theta_{-}}{\theta_{+}}\right)\right\} ,\label{eq:scale factor-1}\\
\phi & =\phi_{0}\left[\left(\theta+\theta_{-}\right)^{2}+\theta_{+}^{2}\right]\ .\label{eq:scalar field-1}
\end{align}

Note that the scalar field dynamics does not depend on $\omega$ anymore and the scale factor has a mild dependence on this parameter. The asymptotic
behavior of the Eq.s~(\ref{eq:scale factor-1})\eqref{eq:scalar field-1} for $\theta\rightarrow\pm\infty$ are
\begin{align}
a\left(\theta\right)&\propto \exp\left(\epsilon\frac{\pi}{2}\sqrt{\frac{2|\omega|}{3}-1}\right)\frac{1}{\theta},\label{eq:scale factor asymp-1}\\
\phi\left(\theta\right)&\propto \theta^2,\label{eq:scalar field asymp-1}
\end{align}
where $\epsilon=\mp1$ for $\theta\rightarrow-\infty$ and $\epsilon=\pm1$ for $\theta\rightarrow+\infty$. In this asymptotic behavior, the cosmic time goes as $t\propto \theta^{-2}$, hence, the scale factor Eq.~(\ref{eq:scale factor asymp-1}) goes as $a\propto t^{\frac{1}{2}}$ in both asymptotic limits $\theta\rightarrow\pm \infty$. Similarly, the asymptotic behavior of the scalar field is $\phi\propto\theta^{2}\propto t^{-1}$. 
The two asymptotic limits $\theta\rightarrow\pm\infty$ describe two distinct possible evolution where the universe behaves as a GR radiation dominated phase. For $\theta\rightarrow-\infty$, the universe starts from an initial singularity at t=0 and expands therefrom with a radiation dominated phase. In the limit $\theta\rightarrow+\infty$, the universe contracts from infinity until it reaches a big crunch singularity again at t=0 during a radiation dominated phase.

The second family is described by $\Delta>0$ $\left(\omega>-\frac{3}{2}\right)$. The general solutions are given by
\begin{align}
a&=a_{0}\left(\theta-\theta_{+}\right)^{\omega/3\Sigma_\mp}\left(\theta-\theta_{-}\right)^{\omega/3\Sigma_\pm}\ ,\label{eq2:scale factor-1}\\
\phi&=\phi_{0}\left(\theta-\theta_{+}\right)^{\left(1\mp\sqrt{{1+2\omega}/{3}}\right)/\Sigma_\mp}\left(\theta-\theta_{-}\right)^{\left(1\pm\sqrt{{1+2\omega}/{3}}\right)/\Sigma_\pm}\ ,\label{eq2:scalar field-1}\\
\Sigma_\pm&=\sigma\pm\sqrt{1+\frac{2\omega}{3}}\label{eq2:A-1}
\end{align}
where now $\theta_{\pm}\equiv(-B\pm\sqrt{\Delta})/2A$  are constants of integration with $\theta_{+}>\theta_{-}$. The solutions for the stiff matter $(\alpha=1)$ reduce to
\begin{align}
a&=a_{0}\left[\theta-\theta_{+}\right]^{\omega/\left[3\left(1\mp\sqrt{1+{2\omega}/{3}}\right)\right]}\left(\theta-\theta_{-}\right)^{\omega/\left[3\left(1\pm\sqrt{{1+2\omega}/{3}}\right)\right]},\label{eq:scale factor-2}\\
\phi & =\phi_{0}\left(\theta-\theta_{+}\right)\left(\theta-\theta_{-}\right)\ .\label{eq:scalar field-2}
\end{align}

The asymptotic behavior of the Eq.s~(\ref{eq:scale factor-2})-\eqref{eq:scalar field-2} show that for $|\theta|\gg \theta_{+}$ we have $a\propto \theta^{-1}\propto t^{\frac{1}{2}}$ and $\phi\propto \theta^2\propto t^{-1}$. Thus, we have the same asymptotic behavior as the $\omega<-\frac{3}{2}$ given by Eq.s~\eqref{eq:scale factor asymp-1}-\eqref{eq:scalar field asymp-1}.

Finally, there is a third family of solutions given by $\Delta=0$ $(\beta_0=1)$ that describes power law solutions, i.e.
\begin{align}
a&=a_{0}\left({\theta}/{\theta_0}\right)^{\sigma/A}
=a_{0}\left({t}/{t_0}\right)^{\sigma/A_\ast}
,\label{eq:scale factor-3}\\
\phi & =\phi_{0}\left({\theta}/{\theta_0}\right)^{\sigma/A}
=\phi_{0}\left({t}/{t_0}\right)^{\left(1-3\alpha\right)/A_\ast}
\ ,\label{eq:scalar field-3}\\
2A_\ast&=4+3\omega\left(1-\alpha^2\right) \ , \label{eq:A-3}
\end{align}
where we have used the relation between the parametric time $\theta$ with the coordinate time $t$. These solutions have singular behavior for $t\rightarrow0$ if the power of the scale factor is positive. This happens when $\omega<-1/(1-\alpha)$ for $\alpha\in \left[-1,1/3\right]$ and $\omega<-4/3(1-\alpha)$ for $\alpha\in \left[1/3,1\right]$.

The third family of solutions can be derived by taking the appropriate limit from the previous two families\footnote{We thank the referees for point out this limiting approach.}. Indeed, for $\Delta<0$ the limit $\Delta\rightarrow0$ implies $\theta_{+}\rightarrow 0$ with $\theta_{-}\neq0$, hence Eq.~\eqref{eq:general solution} shows that $a\propto \left(\theta+\theta_-\right)^{\sigma/A}$. For $\Delta>0$ the limit $\Delta\rightarrow0$ implies $\theta_{+}=\theta_{-}=-B/2A$ and again Eq.~\eqref{eq2:scale factor-1} gives the same result $a\propto \left(\theta- \theta_-\right)^{\sigma/A}$ compatible with Eq.~\eqref{eq:scale factor-3}.

Another interesting asymptotic limit for all these solutions is given by finite time but allowing the BD parameter to increase boundlessly. The limit $\omega\rightarrow\infty$ depend crucially if $\alpha=1$ or not. For $\alpha\neq 1$, the limit $|\omega|\rightarrow\infty$ gives $\sigma =\omega(1-\alpha)$, $A=A_\ast=\frac32\omega(1-\alpha)^2$, hence all three functions diverge such that $\phi \rightarrow \phi_{0}$ in all three cases. Furthermore, the scale factor becomes
\begin{align}
\lim_{\omega\rightarrow-\infty} a & =a_{0}\Big[\left(\theta+\theta_{-}\right)^{2}+\theta_{+}^{2}\Big]^{1/{3(1-\alpha)}} &&,\quad \mbox{for } \beta_0\neq1\ \mbox{and}\ \omega<-\frac32\label{eq:general solution winfty}\\
\lim_{\omega\rightarrow\infty} a & =a_{0}\Big[\left(\theta-\theta_{+}\right)\left(\theta-\theta_{-}\right)\Big]^{1/{3(1-\alpha)}} &&,\quad \mbox{for } \beta_0\neq1\ \mbox{and}\ \omega>-\frac32
\\
\lim_{|\omega|\rightarrow\infty} a & =a_{0}\left({\theta}/{\theta_0}\right)^{2/3\left(1-\alpha\right)}=a_{0}\left({t}/{t_0}\right)^{2/3\left(1-\alpha\right)} &&,\quad \mbox{for}\  \beta_0=1\ \mbox{and }  \omega\neq -\frac32
\end{align}

Therefore, if $\alpha\neq 1$, independently of the $sign(\omega)$, all three families of solutions asymptotically approach GR. However, that is not the case for stiff matter. For $\alpha=1$, in the limit $|\omega|\rightarrow\infty$, $\phi$ does not go to a constant and the scalar field does not go to its GR limit. Indeed, the $\alpha=1$ case has to be studied separately, which is what we shall analyze in the next section.

\section{Exact Power Law Solution}\label{sec:PartSol}

Let us study the dynamic for FLRW universe filled with a stiff matter perfect fluid. Following Gurevich's \emph{et al.} family of solutions, we propose a power law ansatz such that 
\begin{eqnarray}\label{powerlaw}
a=a_{0}t^{r}\quad,\quad\phi=\phi_{0}t^{s}.
\end{eqnarray}
Equating the power in the Klein-Gordon Eq.~\eqref{KG1}, it is easy to check $6r+s=2 $. Furthermore, the coefficients of Eq.s~\eqref{FLRW1}-\eqref{KG1} imply 
\begin{align}
3r^{2} +3rs-\frac{\omega}{2}s^{2}& = \frac{8\pi\rho_{0}}{\phi_{0}}-3 \frac{k}{a_{0}^{2}}t^{2-2r}\, \\
s(s-1)+3rs &= -\frac{16\pi\rho_{0}}{(3+2\omega)\phi_{0}}.
\end{align}

Using the previous relation for $r$ and $s$ and combine these two equations, we find
\begin{eqnarray}
r = \frac{1}{2}\quad, \quad s = -1 \label{basisstiffsolutions}
\end{eqnarray}
for spatial curvature $k=0$, and

\begin{eqnarray}
r = 1 \quad, \quad s = -4.
\end{eqnarray}
for $k\neq0$.
Note also that the equations require the constraint
\begin{eqnarray}\label{phi0rho0wrel}
\phi_{0}=-\frac{32\pi}{(3+2\omega)}\rho_{0}\,
\end{eqnarray}
for flat spatial sections, and
\begin{eqnarray}
\phi_{0} = - \frac{2 \pi}{3+2\omega}\rho_{0}
\end{eqnarray}
with $k=-a_{0}^{2}$ normalized to $-1$ for non-flat. From now on we will focus solely on the solution (\ref{basisstiffsolutions}), for a $k=0$ FLRW universe.

Therefore, in order to have an attractive gravity with positive energy density, i.e. ${\rho_{0}}$ and ${\phi_{0}}>0$, we must have $\omega<-\frac{3}{2}$. During the whole evolution the scale factor mimics a GR radiation dominated expansion, namely $a\propto t^{1/2}$, while the scalar field is always decreasing inversely proportional to the cosmic time $\phi\propto t^{-1}$. There is no GR limit in the sense that the evolution does not depend on the BD parameter $\omega$. Furthermore, Eq.~\eqref{phi0rho0wrel} seems to show that the limit $\omega\rightarrow\infty$ is not even well defined since the gravitation strength is inversely proportional to the scalar field, i.e $G\sim\phi^{-1}$. Notwithstanding, this power law solution is completely consistent for finite values of $\omega$. In the next section we shall study the cosmological perturbations over this particular solution. 

\section{Cosmological Perturbation}\label{sec:Pert}

Consider the background solution found above for a flat FLRW universe filled with stiff matter in BD theory, i.e.
\begin{align}\label{backsol}
a=a_0t^{1/2}\quad  ,\quad \phi=\phi_0t^{-1} \ .
\end{align}

In Ref.~\cite{plinio} the general perturbed equations for a fluid with equation of state of the type $p=\alpha\rho$, with $\alpha$ constant, has been established for the Brans-Dicke cosmology. The full perturbed dynamical system is given by the perturbed version of Eq.s~\eqref{eq:fe-1}-\eqref{eq:fe-2}. However, the evolution of the matter density perturbation can be analyzed using only the perturbed version of the time-time Einstein's equations, the Klein-Gordon, and the conservation of energy-momentum tensor.

The metric perturbation is defined as $h_{\mu\nu}=g_{\mu\nu}-g^{(0)}_{\mu\nu}$, where $g^{(0)}_{\mu\nu}$ is given by the FLRW solution with Eq.~\eqref{backsol} and $k=0$. Following Ref.~\cite{plinio}, we adopt the synchronous gauge where $h_{0\mu}=0$. It is straightforward to calculate the perturbed Ricci tensor, which has time-time component given by
\begin{displaymath}
\delta R_{00}=\frac{1}{a^{2}}\biggr[\ddot{h}_{kk}-2\frac{\dot{a}}{a}\dot{h}_{kk}+2\biggr(\frac{\dot{a}^{2}}{a^{2}}-\frac{\ddot{a}}{a}\biggl)h_{kk}\biggl] \ .
\end{displaymath}
The time-time component of the perturbed energy-momentum tensor and its trace read
\begin{align} \label{21}
\delta T^{00}=\delta\rho  \quad  ,\qquad   \delta T=\delta\rho-3\delta p\ .
\end{align}

Similarly, the perturbation of the scalar field is defined as $\delta \phi(x)=\phi(x)-\phi^{(0)}(t)$. The  d'Alembertian of the scalar field is
\begin{equation}\label{23}
\delta\Box\phi=\delta\ddot\phi+a\dot ah^{kk}\dot\phi-\frac{1}{2a^{2}}\dot{h}_{kk}\dot\phi+3\frac{\dot {a}}{a}\delta\dot\phi-\frac{1}{a^{2}}\nabla^{2}\delta\phi~~.
\end{equation}

It is convenient to define new variables. In particular, we define the usual expression for the density contrast, a similar version for the perturbation of the scalar field, the divergence of the perturbation of the perfect fluid's velocity field $(\delta u^{i})$, and a normalized version of the metric perturbation. They are defined, respectively, as
\begin{eqnarray}\label{defnewvar}
\delta=\frac{\delta\rho}{\rho}\ ,\quad \lambda=\frac{\delta\phi}{\phi}\ ,\quad U=\delta u_{,i}^{i}\ ,\quad h=\frac{h_{kk}}{a^{2}}\ .
\end{eqnarray}

Using \eqref{defnewvar} and decomposing them in Fourier modes $n$, the time-time BD and the Klein-Gordon, equations (\ref{FLRW1}) and (\ref{KG1}), read respectively
\begin{align}
&\frac{\ddot{h}}{2}+H\dot{h}=\frac{8\pi\rho}{\phi}\biggr(\frac{2+\omega+3\alpha(1+\omega)}{3+2\omega}\biggl)(\delta-\lambda)+\ddot{\lambda}+2(1+\omega)\frac{\dot{\phi}}{\phi}\dot{\lambda}\ ,\label{pertFLRW}\\
&\ddot{\lambda}+\biggr(3H+2\frac{\dot{\phi}}{\phi}\biggl)\dot{\lambda}+\biggr[\frac{n^{2}}{a^{2}}+\biggr(\frac{\ddot{\phi}}{\phi}+3H\frac{\dot{\phi}}{\phi}\biggl)\biggl]\lambda=\frac{8\pi(1-3\alpha)\rho}{(3+2\omega)\phi}\delta+\frac{\dot{\phi}}{\phi}\frac{\dot{h}}{2}\ ,\label{pertKG}
\end{align}

In addition, the conservation of energy-momentum tensor decompose in two equations, namely
\begin{align}
&2\dot{\delta}-(1+\alpha)\left(\dot{h}-2U\right)=0\ ,\label{pertconserv1}\\
&(1+\alpha)\left(\dot{U}+(2-3\alpha)H U\right)=\alpha\frac{n^{2}}{a^{2}}\delta\ .\label{pertconserv2}
\end{align}
where again $n$ represents the Fourier mode. In the long wavelength limit, $n\rightarrow0$, Eq.~\eqref{pertconserv2} shows that the perturbation of the four velocity decouples. For a stiff matter fluid, $\alpha=1$, it becomes a growing mode with
\begin{eqnarray}
U \propto a\ .
\end{eqnarray}

This growing mode has nothing to do with BD's extra scalar degree of freedom. Eq.s~\eqref{pertconserv1}-\eqref{pertconserv2} come from perturbing the conservation of the energy-momentum Eq.~\eqref{conservTmunu}, which is identical in GR. For equation of state lower than $\alpha<2/3$, such as for radiation, we have a decaying mode and can ignore it by setting $U=0$ that in turn implies $2\delta=(1+\alpha)h$. In contrast, in our case $\alpha=1$ the growing mode together with Eq.~\eqref{pertconserv1} implies
\begin{eqnarray}\label{dhtheta}
\delta = h - \frac{4}{3}U_0t^{3/2},
\end{eqnarray}
with $U_0$ a constant of integration. For this reason, we will retain this inhomogeneous term.

Using the background expressions (Eq.~\eqref{backsol}), the long wavelength limit $(n\rightarrow0)$, and Eq.~\eqref{dhtheta}, the dynamical system simplifies to 
\begin{align}
\frac{\ddot{h}}{2}
+\frac{\dot{h}}{2t}
+\frac{(5+4\omega)}{4t^2}h
=&\ddot{\lambda}
-\frac{2(1+\omega)\dot{\lambda}}{t}
+\frac{(5+4\omega)}{4t^2}\lambda
+ \frac{(5 + 4\omega)U_0}{3t^{1/2}} \ ,\\
-\frac{\dot{h}}{2t}+\frac{h}{2t^{2}} 
=&
\ddot{\lambda}
-\frac{\dot{\lambda}}{2t}
+\frac{\lambda}{2t^{2}}  
+ \frac{2U_0}{3t^{1/2}}\ .
\end{align}

These equations admit a solution under the form, 
\begin{eqnarray}
h = h_{0}t^{m} + f_0t^{\frac{3}{2}},\quad\lambda=\lambda_{0}t^{m} + g_0t^{\frac{3}{2}},
\end{eqnarray}
with $\lambda_{0}$, $h_{0}$, $f_0$ and $g_0$ are constants. Equating the power in the time parameter and the coefficients of the polynomials, we obtain a set of equations connecting these different constants of integration, namely
\begin{align}
\biggr\{m^2 + \frac{(5 + 4\omega)}{2}\biggl\}h_0 &= 2\biggr\{m^2 - (3 + 2\omega)m + \frac{(5 + 4\omega)}{4}\biggl\}\lambda_0,\\
(1 - m)h_0 &= 2\biggr\{m^2 - \frac{3}{2}m + \frac{1}{2}\biggl\}\lambda_0,\\
f_0 &= - \frac{2}{3}(3 + 4\omega)g_0 + \frac{8}{27}(5 + 4\omega)U_0,\\
f_0 &= - 2g_0 + \frac{8}{3}U_0. 
\end{align}

The constants $h_0$ and $\lambda_0$ give the homogeneous modes, while $f_0$ and $g_0$ give the inhomogeneous modes associated with the growing mode $U$. The homogeneous mode admit four power solution given by $m=0,-1,\frac{1}{2}$ and $1$. The solutions corresponding to $m=0,-1$ are connected with the residual gauge freedom typical of the synchronous gauge. A remarkable novelty is that the physical solutions correspond only to growing modes. In fact, these modes appear also in the long wavelength limit of the radiative cosmological model both in the GR and BD theories \cite{peebles,plinio}. However, there are two interesting aspects connected with these modes: there is no dependence on $\omega$, and when $m = 1$, $h_0 = \lambda_0 = 0$, while for $m = \frac{1}{2}$, $h = 0$ and $\lambda_0$ is arbitrary. Thus, the most important perturbative modes are the inhomogeneous modes, which are represented by $h=f_0t^{3/2}$, $\lambda=g_0 t^{3/2}$, with
\begin{eqnarray}
f_0 &=& \frac{2}{9}\frac{(7 + 2\omega)}{\omega}U_0\ ,\\
g_0 &=& - \frac{4}{9}\frac{(7 + 6\omega)}{\omega}U_0\ .
\end{eqnarray}

An important feature of these inhomogeneous solutions is that the perturbation grows very quickly with the scale factor, $\delta \propto a^3$. It is also interesting to contrast with the same situation in GR where this inhomogeneous modes identically cancel for the pure stiff matter case~\cite{amendola2015,weinberg2008}. As a final remark, we should stress that these homogeneous and inhomogeneous modes have a well defined $\omega\rightarrow\infty$ limit. However, these solutions depend on the background solution, which is inconsistent in this limit.

\section{Dynamical system analysis}\label{sec:DynSys}

The power law solution for stiff matter fluid in BD displayed in section~\ref{sec:PartSol} has distinct features compared with Gurevich's families with $\alpha\neq1$ see section~\ref{sec:GureSol}. In order to compare these solutions, we perform a dynamical system analysis. Instead of using the conservation equation~\eqref{conservTmunu}, in this section we shall use BD and Klein-Gordon equations of motion. For a flat FLRW, the dynamical system reads
\begin{align}
\left(\frac{\dot{a}}{a}\right)^{2}+\left(\frac{\dot{a}}{a}\right)\frac{\dot{\phi}}{\phi} = & \frac{8\pi\rho}{3\phi}+\frac{\omega}{6}\left(\frac{\dot{\phi}}{\phi}\right)^{2}\ ,\label{dynsysH00}\\
\left(\frac{\ddot{a}}{a}\right)-\left(\frac{\dot{a}}{a}\right)\frac{\dot{\phi}}{\phi} =&  -\frac{8\pi}{3\phi}\frac{(3+\omega)\rho+3\omega p}{2\omega+3} -\frac{\omega}{3}\left(\frac{\dot{\phi}}{\phi}\right)^{2} ,\label{dynsysHii}\\
\frac{\ddot{\phi}}{\phi}+3\left(\frac{\dot{a}}{a}\right)\frac{\dot{\phi}}{\phi}=&\frac{8\pi}{(3+2\omega)}\frac{\left(\rho-3p\right)}{\phi}\ .\label{dynsysKG}
\end{align}

It is convenient to define the Hubble factor $H={\dot{a}}/{a}$ and its analogous for the scalar field, namely $F={\dot{\phi}}/{\phi}$. Restricting ourselves to the stiff matter $(p=\rho)$, we can combine the hamiltonian constraint Eq.~\eqref{dynsysH00} with Eq.~\eqref{dynsysHii} and \eqref{dynsysKG} obtaining the following dynamical autonomous system: 
\begin{align}
\dot{H} & =  -\frac{6}{(3+2\omega)}\left((1+\omega)H^{2}+\frac{\omega}{3} HF+\frac{\omega}{12} F^{2}
\right)\ , \label{Hponto}\\
\dot{F} & = 
 -\frac{6}{(3+2\omega)}\left(H^{2}+\frac{(5+2\omega)}{2}HF+\frac{(3+\omega)}{6}F^2 \right) \ , \label{Fponto}
\end{align}

In fact, note that Eq.s~\eqref{Hponto}-\eqref{Fponto} can simultaneously describe the stiff matter and the vacuum $(p=\rho=0)$. It is easy to check that there are two fixed points for this dynamical system
\begin{align}
H&=F=0  \quad \mbox{corresponding to the Minkowski case,}\\
H&=-\frac{F}{3} \quad \mbox{with}\quad  \omega=-\frac{4}{3}\ .
\end{align}

We can also find the invariant rays defined by the condition $F=q H$  with $q$ constant, which correspond to power law solutions of the system. Using Eq.~\eqref{powerlaw}, this condition translates into $q={s}/{r}$, where $r$ and $s$ are the powers in time of the scale factor and of the scalar field, respectively . Imposing this condition and combining the resulting expressions we find the following third order polynomial for $q$
\begin{eqnarray}
(q+2)\biggr(\frac{\omega}{2}q^{2}-3q-3\biggl)=0,
\end{eqnarray}
with solutions given by
\begin{equation}
q =  -2  \quad ,\quad 
q_{\pm}  =  \frac{3}{\omega}\biggr(1\pm\sqrt{1+\frac{2}{3}\omega}\biggl)
 \ .\label{qstiff}
\end{equation}

The first root corresponds to the power law solution found previously, for which gravity is attractive only if $\omega<-{3}/{2}$. Indeed, using $F=qH$, the constraint Eq.~\eqref{dynsysH00} reads
\begin{eqnarray}\label{constq}
\biggr(-\frac{\omega}{2}q^{2}+3q+3\biggl)H^{2}=\frac{8\pi\rho}{\phi}\ .
\end{eqnarray}
One can immediately see that if $q=-2$  then the energy density is positive only for $2\omega+3<0$ as already argued in Eq.~\eqref{phi0rho0wrel}. The other two roots correspond to the vacuum solution. Again, Eq.~\eqref{constq} shows that for $q=q_\pm$ the left hand-side of the above equation vanishes implying that $\rho=0$. Note also that the invariant rays $q=q_\pm$ disappear when $\omega<-\frac{3}{2}$ (the roots become imaginary). Varying $\omega$ into negative values makes the two $q_\pm$ rays collapse into $q=-2$ when $\omega=-3/2$. For $\omega<-3/2$ only the $q=-2$ invariant ray remains (see Fig.~\ref{phase}).

The $q=-2$ invariant ray does not depend on $\omega$ which means that is insensitive to the $\omega\rightarrow\infty$ limit. On the other hand, the $q=q_\pm$ decays as
\begin{equation}
\lim_{\omega\rightarrow\infty} q_\pm=\pm\sqrt{\frac{6}{\omega}}\ .
\end{equation}

Thus, since $F=q H$, for arbitrary finite values of $H$ we have $\dot{\phi}\rightarrow 0$ in the $\omega\rightarrow\infty$ limit. Naively, one could expect that a vacuum solution with $\dot{\phi}\rightarrow 0$ should approach the Minkowski spacetime. However, the term $\omega\, \dot{\phi}^2$ does not go to zero in this limit producing a power law expansion with $a\propto t^{1/3}$. Indeed, it has been shown in Ref.~\cite{brando2018} that in this regime the $\omega\, \dot{\phi}^2$ term behaves as an effective stiff matter like term, which is responsible for the $a\propto t^{1/3}$ evolution of the scale factor.

\subsection{Singular Points at Infinity}

The stability of the fixed point at the origin of the phase space can be inferred directly from the phase space diagrams. However, the stability of the invariant rays of the dynamical system must be analyzed at infinity. For this purpose we use the Poincar\'{e} central projection method~ \cite{sansoniconti,bogo,juliojoel}, using the coordinate transformation
\begin{equation}\label{transf}
H = \frac{h}{z}, \ \ F = \frac{f}{z}, \quad \quad \mbox{with}\ h^2 + f^2 + z^2 =1\ .
\end{equation}
Eq.s~\eqref{Hponto}--\eqref{Fponto}) can be recast as $P(H,F)dF - Q(H,F)dH = 0$, which combined with equation Eq.~\eqref{transf} gives us
\begin{equation}\label{dfdhdz}
-zQdh + zPdf + \left(hQ-fP\right)dz=0\ .
\end{equation}
where the functions $P(h,f,z)$ and $Q(h,f,z)$ are given by
\begin{align}
P(h,f,z) &= -\frac{6}{(3+2\omega)}\left((1+\omega)h^{2}+\frac{\omega}{3} hf+\frac{\omega}{12} f^{2}\right)\ , \\
Q(h,f,z) &= -\frac{6}{(3+2\omega)}\left(h^{2}+\frac{(5+2\omega)}{2} hf+\frac{(3+\omega)}{6}  f^{2}\right)\ .
\end{align}
Collecting all term we can explicitly write Eq.~\eqref{dfdhdz} in term of the projective coordinates as
\begin{align}\label{dfdhdzexpl}
&z\Big[ 6h^2 +3(2\omega+5)hf + (3+\omega)f^2 \Big]dh\nonumber \\
-&z \left[6(1+\omega)h^2 + 2\omega hf + \frac{\omega}{2} f^2 \right]df\nonumber \\ 
-& \left[ 6 h^3 - \frac{\omega}{2} f^3 + (3-\omega)hf^2 + 9 h^2f \right] dz= 0\ .
\end{align}

The singular points at infinity have projective coordinates in the plane $(h,f,z=0)$. Given Eq.s~\eqref{transf} and \eqref{dfdhdzexpl}, they are solutions of the system:
\begin{equation}\label{hfsist}
\begin{split}
&h^2+f^2=1\ , \\
& 6 h^3 - \frac{\omega}{2} f^3 + (3-\omega)hf^2 + 9 h^2f =0\ .
\end{split}
\end{equation}

In order to find the invariant rays, we substitute ${f}/{h} = q$ in the system Eq.s~(\ref{hfsist}). As expected, there are three invariant rays
\begin{align}
q&=-2 \ :
&h&= \pm \frac{1}{\sqrt{5}} \ ,
&&f=qh \ , \label{ray1}\\
q_{+}&=\frac{3}{\omega}\left(1+\sqrt{1+\frac{2}{3}\omega}\right) \ :
&h&= \frac{1}{\sqrt{1+q_{+}^2}} \ ,
&&f=q_{+}h\ ,   \label{ray2}\\
q_{-}&= \frac{3}{\omega} \left(1-\sqrt{1+\frac{2}{3}\omega}\right)\ :
&h&= \frac{1}{\sqrt{1+q_{-}^2}}\ ,
&& f=q_{-}h\ .\label{ray3}
\end{align}

The analytical expressions of the solutions of the scale factor and the scalar field that correspond to these rays are given by\\
For $q=-2$ :
\begin{equation}
a(t) \propto t^{1/2}\quad  ,\quad  \phi(t) \propto t^{-1} \ , \label{invraystiff}
\end{equation}
For $q_{+}= \frac{3}{\omega}\left(1+\sqrt{1+\frac{2}{3}\omega}\right)$ :
\begin{equation}
a(t) \propto t^{{\omega\left(3+q_{+}\right)}/{3(4+3\omega)}}\quad  , \quad \phi(t) \propto t^{{\left(4-\omega q_{+}\right)}/{(4+3\omega)}}\ ,
\label{invrayvac1}
\end{equation}
For $q_{-}= \frac{3}{\omega}\left(1-\sqrt{1+\frac{2}{3}\omega}\right)$ :
\begin{equation}
a(t) \propto t^{{\omega\left(3+q_{-}\right)}/{3(4+3\omega)}}\quad  , \quad  \phi(t) \propto t^{{\left(4-\omega q_{-}\right)}/{(4+3\omega)}}\ .
\label{invrayvac2}
\end{equation}

\begin{figure}[h]
\centering
\includegraphics[width=0.27\textwidth]{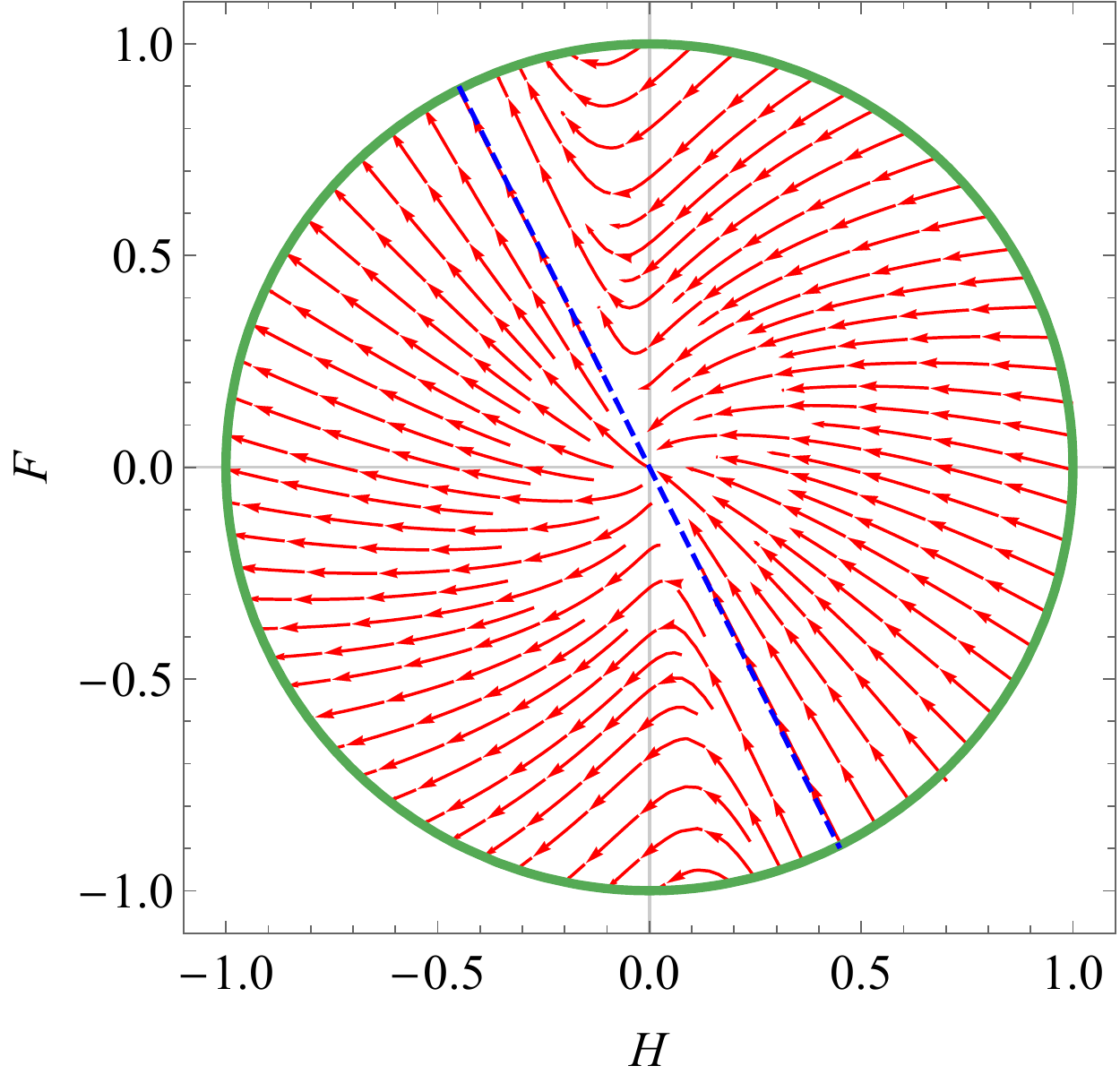}
\includegraphics[width=0.27\textwidth]{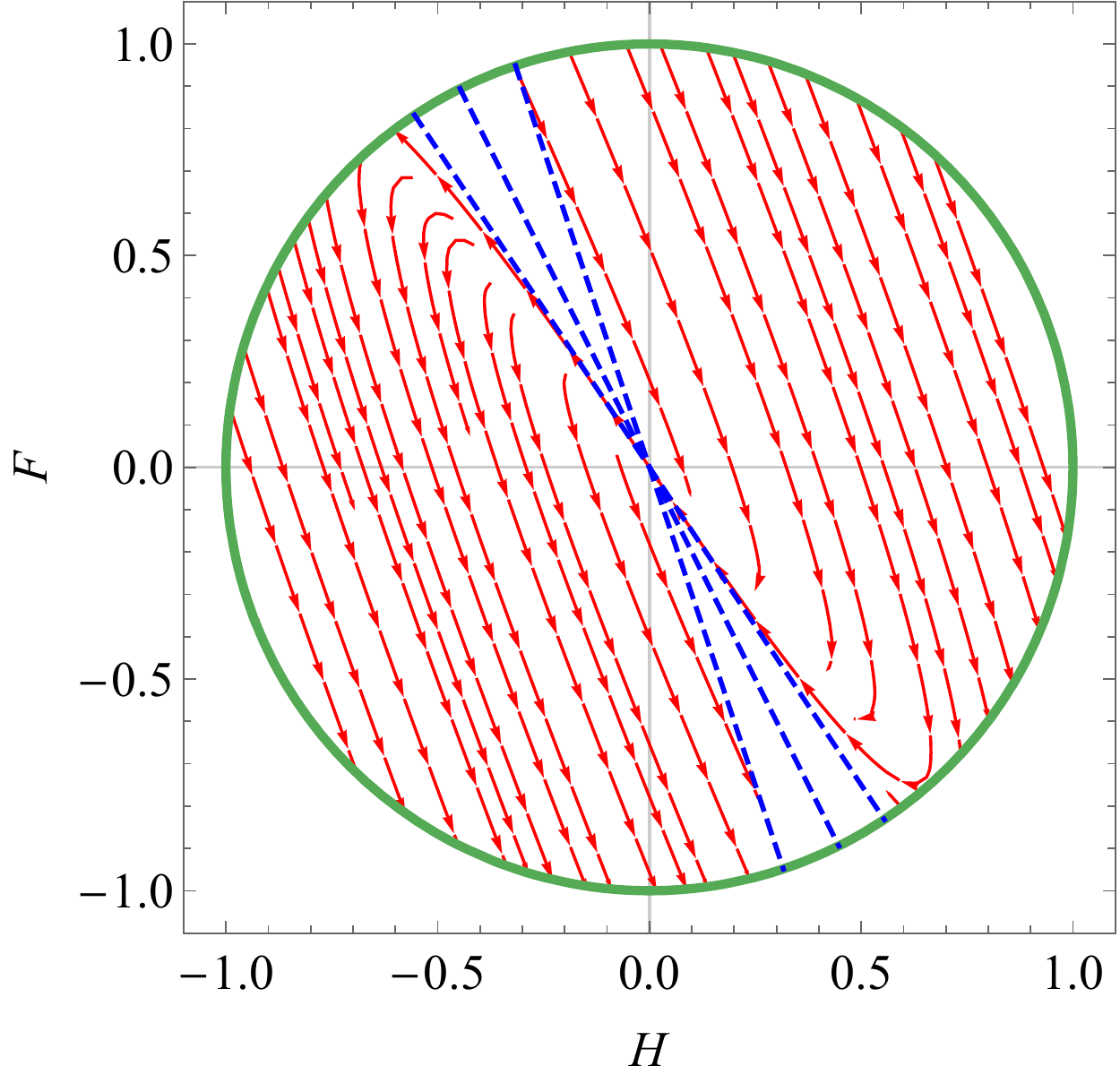}
\includegraphics[width=0.27\textwidth]{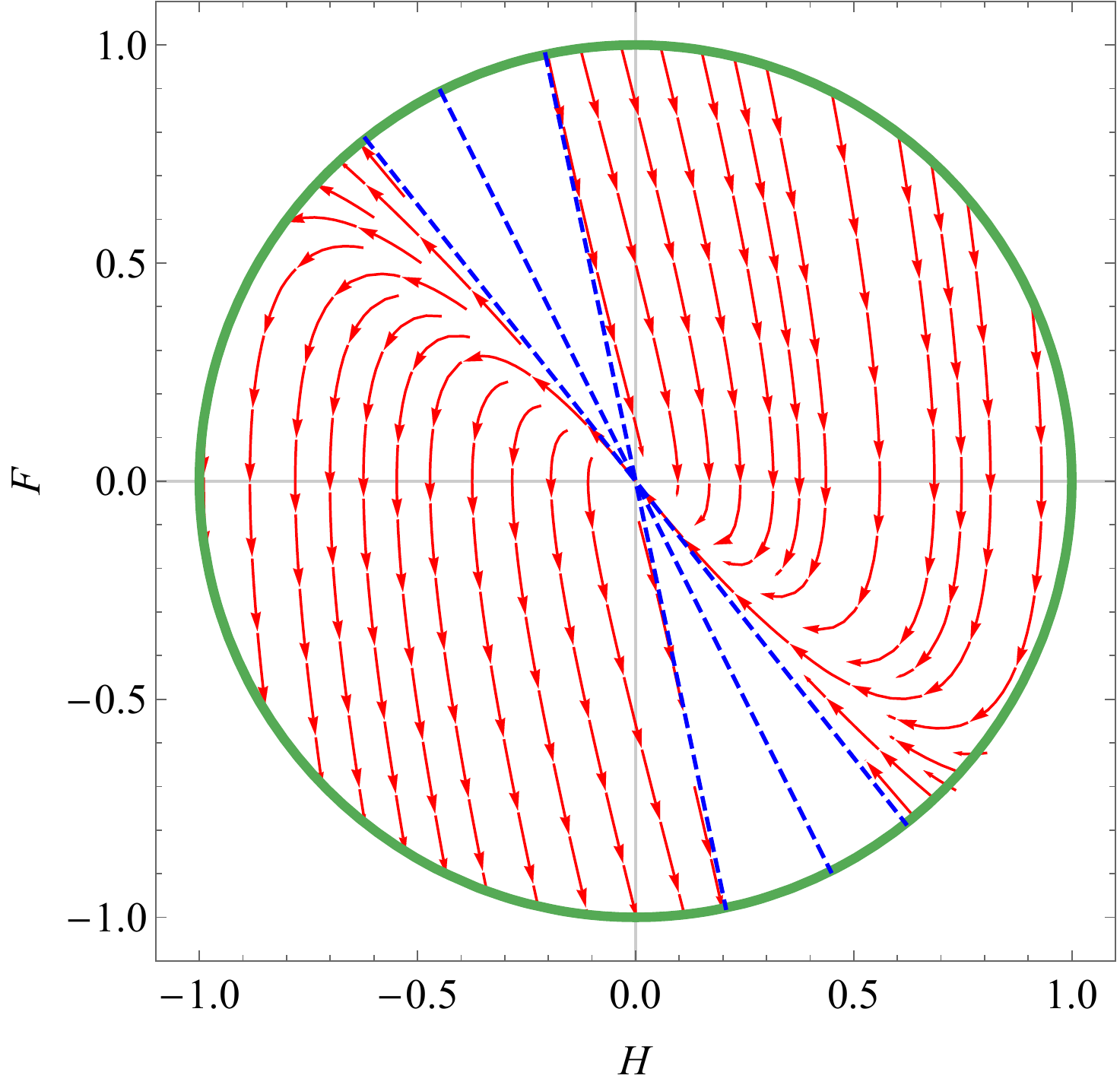}\\
\includegraphics[width=0.27\textwidth]{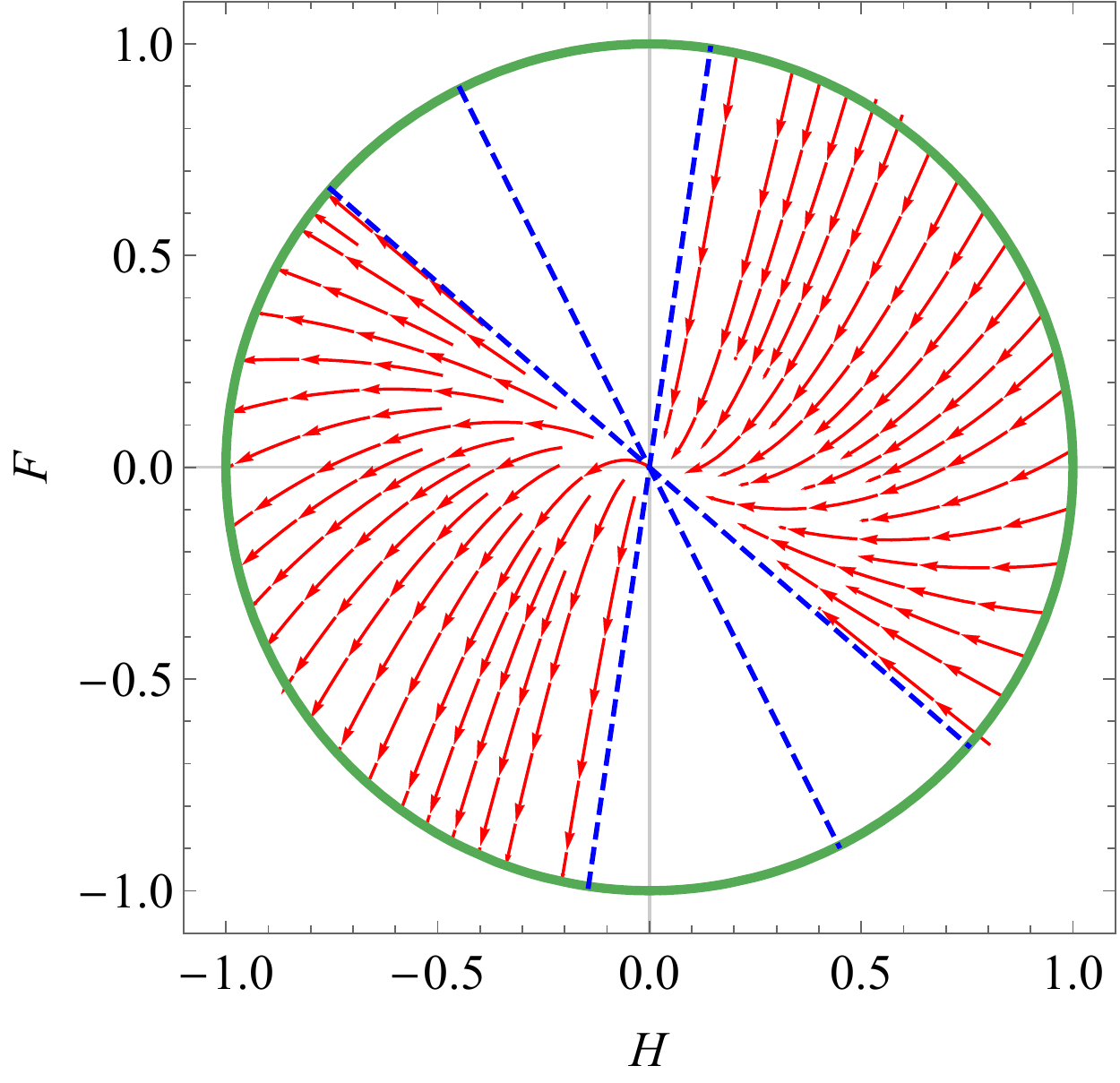}
\includegraphics[width=0.27\textwidth]{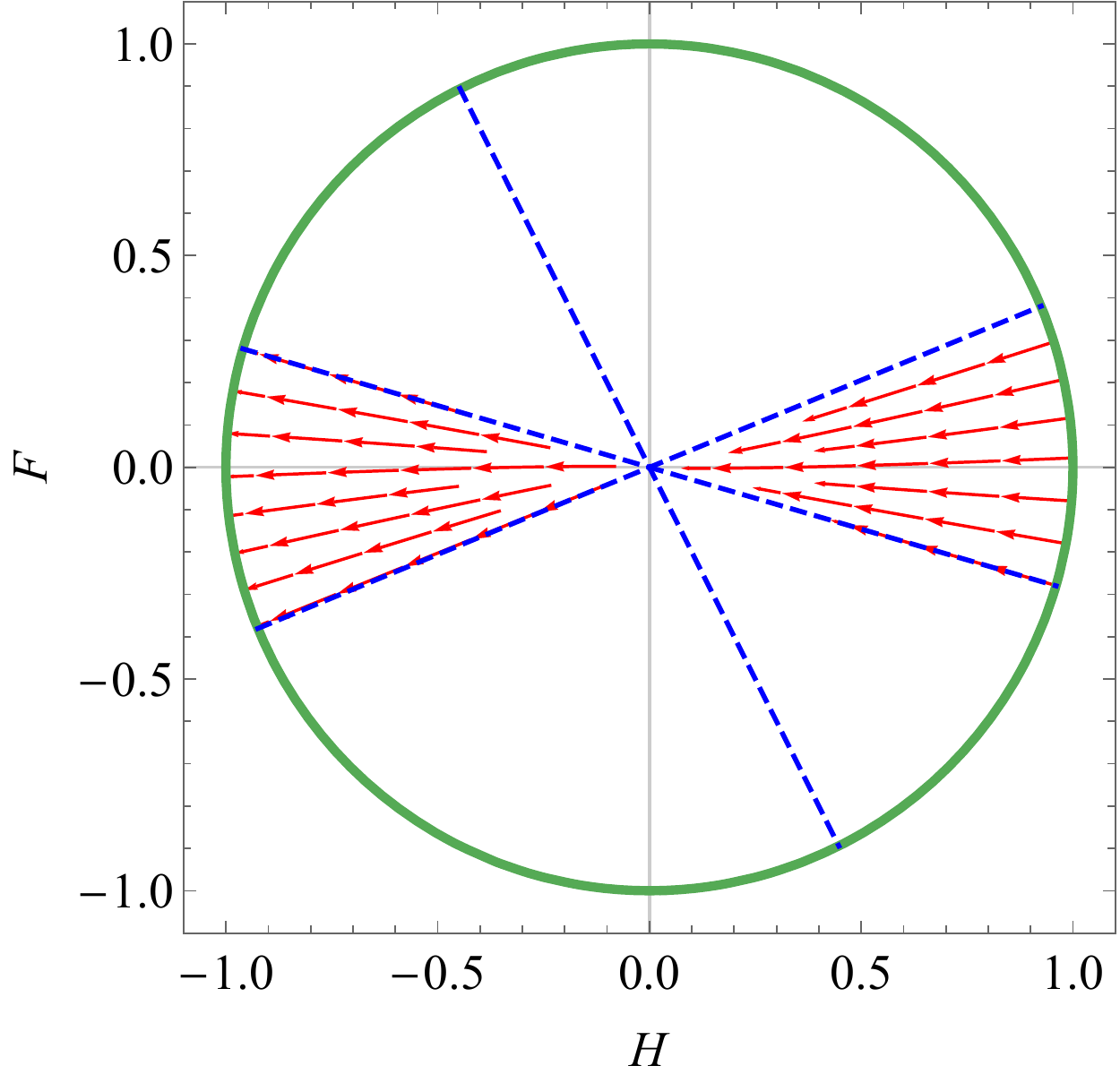}
\includegraphics[width=0.27\textwidth]{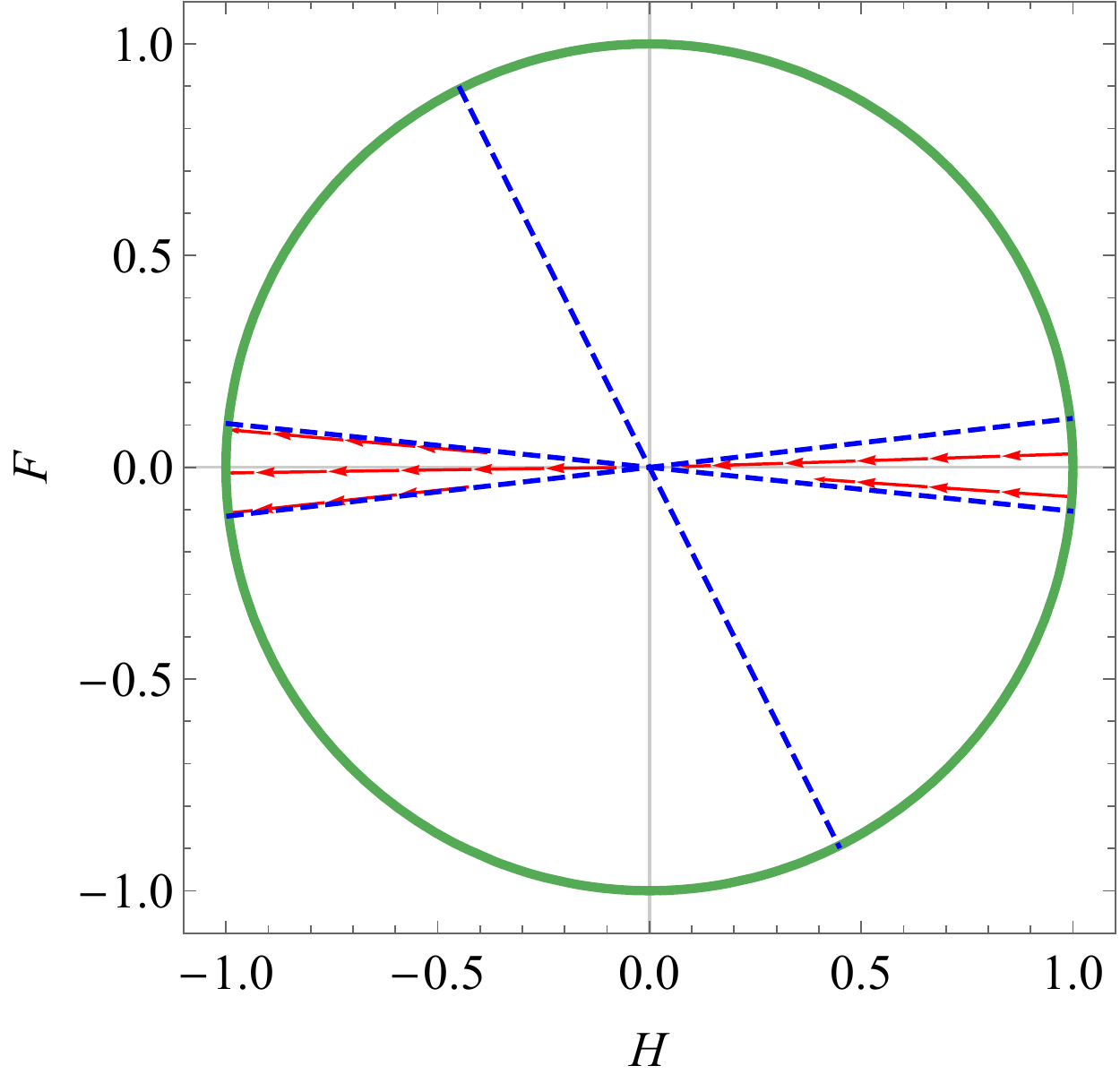}
\caption{Compacted phase portraits using the variables $h$, $f$ that are respectively related to $H$ and $F$ by Eq.~\eqref{transf}. Each portrait uses a different value of BD parameter $\omega$. In particular, we used the values $\omega=-5,-4/3,-1,1,50,500$ respectively from the top left to the bottom right. The unity circle corresponds to the projected singular points at infinity. The dashed straight lines depict the invariant rays Eq.s~(\ref{ray1}-\ref{ray3}) and the empty region has been excluded since it corresponds to the unphysical situation of negative values of the energy density.}
\label{phase}
\end{figure}

The phase portrait for six different values of $\omega$: $-5$, $-4/3$, $-1$, $1$, $50$ and $500$ are plotted in Fig.~\ref{phase}. As mentioned before, the two spread invariant rays for $\omega>-3/2$ are related to the $q_\pm$ vacuum solution ($\rho=0$), while the invariant ray in the middle is for $q=-2$, which corresponds to our solution Eq.~\eqref{powerlaw} for $\omega<-3/2$. Increasing the value of $\omega$ makes the $q_\pm$ invariant rays to move away from the $q=-2$ invariant ray. As can be seen by Eq.~\eqref{constq}, the region between the two invariant rays $q_\pm$ corresponds to negative values of the energy density, hence should be excluded on physical basis. Additionally, for large values of $\omega$ the $q_\pm$ invariant rays tend to lay along the $F=0$ line, which represent the $\omega\rightarrow \infty$ limit.

\section{Discussion}\label{sec:Concl}

In this paper we analyzed the cosmological solution for a perfect fluid in Brans-Dicke theory and showed that stiff matter $(p=\rho)$ is a very particular solution. There are power law solution with the scale factor and the scalar field, respectively, proportional to $a\propto t^{1/2}$ and $\phi\propto t^{-1}$. Even though the matter content behaves as stiff matter, the cosmological evolution mimics a radiation dominated dynamics in General Relativity. Furthermore, the scalar field gives the effective gravitational strength, hence, gravity becomes stronger with the expansion of the universe.

Eq.~\eqref{phi0rho0wrel} shows that the scalar field is inversely proportional to the BD parameter $\omega$. This condition is commonly understood as a sufficient condition for a well defined GR limit. However, we have shown that this is not the case for the power law solution \eqref{powerlaw}. 

The scalar cosmological perturbation also has interesting features. The velocity field for the stiff matter fluid has a growing mode $U$ that is proportional to scale factor. This extra contribution produce new polynomials solutions for the density contrast $\delta=\delta\rho/\rho$, the fractional scalar field perturbation $\lambda=\delta \phi/\phi$ and the tensor perturbation $h=h_{kk}/a^2$. The homogeneous mode has four power solution in cosmic time $t^m$ with $m=-1,0,1/2,1$. The first two are connected with the residual freedom of the synchronous gauge and the other two are the physical solutions corresponding to two growing modes. There is no decaying mode. The inhomogeneous mode related to the growing mode $U$ goes as $t^{3/2}\propto a^3$, hence it is a steep growth if compared with the standard cosmological model.

The dynamical systems analysis developed in section~\ref{sec:DynSys} shows the existence of three invariant rays associated with the system Eq.s~\eqref{Hponto}-\eqref{Fponto}. The first correspond to the power law solution of section~\ref{sec:PartSol} with constant of proportionally $q=-2$. The other two are vacuum solutions with constant of proportionally $q_\pm$ given by Eq.~\eqref{qstiff}. For $\omega<-3/2$ there is only one invariant ray associated with $q=-2$ while for $\omega>-3/2$ there are two vacuum invariant rays $q_\pm$. The region in the phase space diagram between these two rays has negative energy density, which is unphysical. Increasing the value of the BD parameter $\omega$, these two rays rotate away from each other expanding the unphysical region. The limiting case is for $\omega\rightarrow\infty$ when both $q_\pm$ tends to zero. This limit has a vacuum solution with constant scalar field $F=0$ but the scale factor increases as $a\propto t^{1/3}$, which is characteristic of a FLRW universe with stiff matter in GR. Even though this is the vacuum case and the scalar field is constant, we do not approach Minkowski spacetime. 

We would like to point out to a possible cosmological realization of our solution (\ref{basisstiffsolutions}). Alternative theories of gravity, such as BD, are commonly used in cosmology to explain the late time acceleration of our universe. One of these models are Quintessence models, in which the matter component is described by a minimally coupled scalar field, with a potential $V(\psi)$. Some of these models~\cite{amendola2015}, are such that its potential goes to zero at early times, which implies that in this period the scalar field has a stiff matter type equation of state. Therefore, the model described in this work can be interpreted as the early time description of such models.

It is well known in the literature that there are examples where the BD parameter scales as $\phi\sim 1/\sqrt{\omega}$ but the system does not approach a GR regime in the limit $\omega\rightarrow\infty$. Nevertheless, it is commonly expected to recover GR in this limit if $\phi\sim 1/\omega$ and the matter energy-momentum tensor has a nonzero trace. We have explicit showed an exact BD solution with $\phi\sim 1/\omega$ and $T^\mu{}_\mu\neq 0$ that does not approach GR in the limit $\omega\rightarrow\infty$.

\section*{Acknowledgments}
The authors are grateful to Luca Amendola for valuable comments, and to an anonymous referee for important corrections. The authors would also like to thank and acknowledge financial support from the National Scientific and Technological Research Council (CNPq, Brazil) the State Scientific and Innovation Funding Agency of Esp\'\i rito Santo (FAPES, Brazil) and the Brazilian Federal Agency for Support and Evaluation of Graduate Education (CAPES, Brazil).

\end{document}